# K-shell Core Electron Excitations in Electronic Stopping of Protons in Water from First Principles

*Yi Yao, Dillon C. Yost, and Yosuke Kanai*[*]

Department of Chemistry, University of North Carolina at Chapel Hill, North Carolina, 27599, USA

Understanding the role of core electron excitation in liquid water under proton irradiation has become important due to the growing use of proton beams in radiation oncology. Using a first-principles, non-equilibrium simulation approach based on real-time time-dependent density functional theory, we determine the electronic stopping power, the velocity-dependent energy transfer rate from irradiating ions to electrons. The electronic stopping power curve agrees quantitatively with experimental data over the velocity range available. At the same time, significant differences are observed between our first-principles result and commonly-used perturbation theoretic models. Excitations of the water molecules' oxygen core electrons are a crucial factor in determining the electronic stopping power curve beyond its maximum. The core electron contribution is responsible for as much as one-third of the stopping power at the high proton velocity of 8.0 a.u. (1.6 MeV). K-shell core electron excitations not only provide an additional channel for the energy transfer but they also significantly influence the valence electron excitations. In the excitation process, generated holes remain highly localized within a few angstroms around the irradiating proton path whereas electrons are excited away from the path. In spite of its great contribution to the stopping power, K-shell electrons play a rather minor role in terms of the excitation density; only 1% of the hole population comprises K-shell holes even at the high proton velocity of 8.0 a.u.. The excitation behavior revealed here is distinctly different from that of photon-based ionizing radiation such as X/γ-rays.

When a highly energetic ion travels through and interacts with matter, its kinetic energy is transferred into the target material's electronic and nuclear subsystems. This energy loss of the projectile ion can arise from both elastic collisions with nuclei (nuclear stopping) and inelastic scattering events (electronic stopping). When the particle's kinetic energy is sufficiently large (on the order of ~10 keV per nucleon), the major contribution to the energy transfer comprises electronic stopping wherein the projectile ion induces massive electronic excitations in the target matter[1-2]. This electronic stopping phenomenon is at the heart of emerging ion bean cancer therapies. The use of proton beam radiation over more conventional radiation based on X/γ-ray photons is often considered more effective because of the ion's distinct spatial energy deposition profile with a very sharp peak.[3-4] By calibrating the initial kinetic energy of the protons, this energy deposition peak can be tuned to coincide with the location of the tumour. This energy deposition profile is largely determined by electronic stopping power, which measures the rate of energy transfer from the charged particle to electrons in matter per unit distance of the energetic particle's movement.[1,5-7] The stopping power is a continuous function of the particle velocity, and the velocities near the maximum of the stopping power are responsible for the formation of the sharp energy deposition peak for ions like protons. Because liquid water makes up the majority of matter in human cells, various models have been developed for the electronic stopping power in liquid water over the years[8-16] including our earlier first-principles theory result[17-18]. At the same time, only limited experimental measurements exist near the stopping power maximum, and various theoretical models are currently used with empirically fitted parameters. Furthermore, unraveling the details of the excitation behavior in the electronic stopping process has become important. Proton radiation is generally considered as being similar to other types of ionizing radiation like X/γ-ray photons, which undergo Compton scattering and also core electron excitation. However, the extent to which proton radiation excites valence and core electrons is not understood. Indeed, this is complicated by the fact that the ratio of valence to core electron excitations depends on the irradiating proton velocity. In radiation oncology, an empirical factor such as relative biological effectiveness is used to take into account differences between the proton radiation and X-ray photon radiation for convenience, but many now call for a better mechanistic understanding of the radiation at the molecular level[19]. In this Letter, we discuss the role of K-shell core electron excitations in liquid water under proton irradiation by accurately determining the electronic stopping power and simulating

---

[*] E-mail: ykanai@unc.edu

quantum dynamics of electronic excitations from first principles.

We apply our recently developed non-equilibrium dynamics simulation approach based on real-time time-dependent density functional theory (RT-TDDFT)[17-18, 20-23] to simulate the non-perturbative response of the electronic system to a fast-moving projectile proton. In this approach, the electronic stopping power can be obtained from the rate of electronic energy change at different projectile proton velocities as discussed in our earlier work[21,24]. We use our implementation of RT-TDDFT based on a planewave-pseudopotential (PW-PP) formalism[20, 25] in the Qbox/Qb@ll code[26-27]. Simulating the 1s core (i.e. K-shell) electron excitations of oxygen atoms in this study requires us to go beyond several standard approximations typically used in the PW-PP formalism. The oxygen and hydrogen atoms in liquid water are described by all-electron pseudopotentials that are generated using the optimized Norm-Conserving Vanderbilt scheme[28-29], for which multiple projectors are used for the explicit treatment of the 1s electrons of oxygen atoms in the simulation. The validity of the all-electron pseudopotentials was checked by calculating the core-level optical excitation spectrum of a single water molecule as shown in the Supplemental Material. Unlike previous RT-TDDFT studies of electronic stopping in which pseudopotentials are used for the projectile proton[17-18], here we use a bare Coulomb potential for representing the proton because an accurate description of the K-shell core excitations is necessary, especially for large proton velocities (see Supplemental Material for details). Consequently, the use of a planewave kinetic energy cutoff of up to 250 Ry for expanding the Kohn-Sham wavefunctions was required, and an extrapolation was used for calculating the stopping power at high velocities (see Supplemental Material for details). We employed the PBE GGA approximation[30] for the exchange-correlation potential because we found that the use of the more advanced SCAN meta-GGA does not change the results[31-33] (Supplemental Material Fig. S7). The liquid water structure was generated by taking a snapshot after preforming a 10 ns classical molecular dynamics simulation at 300K with SPC/E force field[34]. Our simulation cell contains 38 water molecules with periodic boundary conditions (8Å × 8Å × 17.73Å), and the projectile proton travels in the +z direction. This simulation was compared to a larger simulation cell with 170 water molecules (12Å × 12Å × 35.45Å), and no appreciable finite size errors were found. In order to determine electronic stopping power accurately using the non-equilibrium simulation approach, an ensemble average of projectile proton trajectories is necessary[23]. 64 proton projectile trajectories (paths) were sampled evenly on a grid dividing the cross section of the xy simulation cell plane. In total, 64 independent RT-TDDFT simulations were performed for each velocity. The convergence of this sampling was confirmed by comparing to a more extensive sampling that includes 256 paths. Albeit computationally expensive, this trajectory sampling ensures that the ensemble average contains projectile proton trajectories that cover a wide range of impact parameters with respect to the atoms in the target matter, which is especially important when core electrons are excited[23,35]. The error bars on the stopping power reported here are the standard error of the mean calculated based on these 64 paths. Because the K-shell core electron excitation is found to be important in the high velocity regime, we also verified that close/small impact parameters are accurately sampled. These technical, but important, details are discussed in the Supplemental Material, in addition to comparisons to our earlier work[18], which did not consider core electron effects.

The calculated stopping power as a function of the proton velocity ranging from 0.5 to 8 a.u. (corresponding to the kinetic energy of 6.2 keV-1.6 MeV) is compared to the available experimental stopping power data[36-37] and to the so-called SRIM[16] model in Figure 1. The only experimental data available in this velocity range are the measurements by Shimizu, et al.[36-37]. We note, for completeness, that the reliability of this measurement has been questioned on the basis of the Bethe model[38]. The SRIM model is based on extending the Lindhard-Scharff-Schiott theory[39] with inputs from available experiments, and it is widely used as a standard reference. Though there is no reported experimental data for velocities less than 3.5 a.u. for liquid water, the SRIM result relies on existing experimental data of solid water (ice) to estimate the stopping power of liquid water. Our first-principles result is in excellent agreement with these two references. The peak of our calculated stopping power (i.e. the Bragg peak) is at v=1.73 a.u., and the stopping power of 0.165±0.010 a.u. agrees well with the SRIM model which shows the Bragg peak at v=1.72 a.u. and stopping power of 0.165 a.u. at this velocity. For comparison, we also show the seminal Bethe model[40] with mean excitation energy parameter of I=78eV as recommended by the International Commission on Radiation Units and Measurement[8] and one of the most recent models by Emfietzoglou and co-workers[10-12] based on perturbation theory. For the Bethe model, the Bragg peak lies around v=1.98 a.u. with a corresponding stopping power of 0.160 a.u.. As widely recognized, the Bethe model significantly underestimates the stopping power for low projectile velocities, and it does not obey the correct linear dependence around zero velocity[41]. At the same time, the Bethe model is remarkable in that the model correctly captures the stopping power behavior for the large projectile velocities beyond the peak velocity with only a single parameter to account for the target matter, the mean excitation energy. The Emfietzoglou's model goes beyond the Bethe model, and it tends to the correct limits in both low and high velocities. However, Emfietzoglou's model shows the Bragg peak at around v=1.80 a.u. with the stopping power of 0.130 a.u., which significantly underestimates the magnitude of the electronic stopping power with respect to our first-principles result.

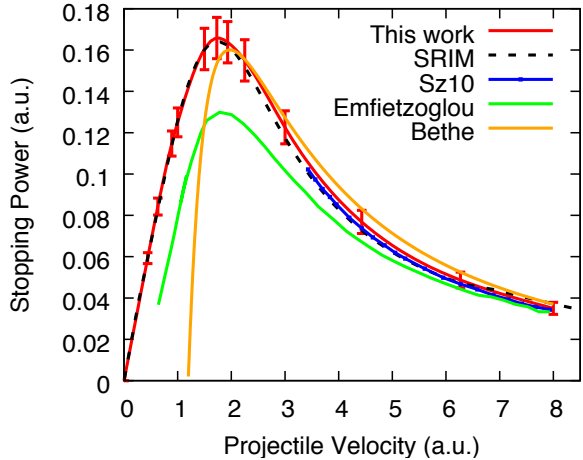

FIG 1 Electronic stopping power curve from our first-principles simulation, in comparison to the experimental data by Shimizu et al.[36-37] (Sz10), SRIM[16] model, the Bethe model[40] with I=78 eV recommended by International Commission on Radiation Units and Measurement[8], and the Emfietzoglou's model[10-12].

One of the most pressing challenges is to elucidate the importance of the K-shell core electron excitations. Widely used in radiation oncology, X/γ-ray radiation could effectively excite deep core electrons, undergoing Auger effect[42]. Empirical models have indicated that proton radiation does not excite K-shell core electrons appreciably for the proton velocities near the Bragg peak but only for much large velocities[11]. In recent years, differences between X/γ-ray and proton radiation have been examined more carefully in the radiation therapy literature[19]. However, our understanding of proton radiation is still quite limited, even for such an important biological matters like liquid water. Here, we examine the extent to which the K-shell core electron excitations play a role in the electronic stopping of protons in liquid water. In the literature, a separate K-shell contribution to stopping power is widely used, as in the Emfietzoglou's model[12]. However, in addition to providing an extra channel for the energy transfer from the projectile proton, electronic excitations of K-shell core electrons also influence the valence electron excitations. This is commonly known as "shake-up" effect[43] in the related context of X-ray absorption. In reality, it is therefore not possible to separate the electronic stopping power in terms of contributions from the valence electrons and core electrons independently as is widely done in empirical models[11-12,44-47]. Using first-principles theory, we can quantify how much the stopping power is influenced by the presence of the K-shell core electrons by calculating the stopping power with and without including the core electrons in our simulations as shown in the top panel of Figure 2. For convenience, we refer to the difference in these two stopping power curves as $\Delta S^{core}$. The valence electron contribution indeed accounts for >99% of the stopping power for velocities less than 1.5 a.u. However, for the velocities larger than 1.5 a.u., the K-shell stopping power contribution, $\Delta S^{core}$, starts to increase, from 0.002 a.u. (2% of the stopping power) at v=1.73 a.u. to 0.012 a.u. (25% of the stopping power) at v=6.27 a.u.. For the highest velocity of 8.0 a.u. we considered here, the stopping power is 28% higher when the core electrons are present. This observation differs significantly from the estimated K-shell electron contribution based on various empirical models (Emfietzoglou/Drude[11-12], and Hydrogenic generalized oscillator strength[11, 45, 48-49]) as shown in the bottom panel of Figure 2. For instance, the Emfietzoglou's model[12] predicts that the K-shell contribution starts to become important only at much greater velocities of > 3.5 a.u. (Figure 2 (Bottom)), and K-shell core electrons are responsible for less than 10% of the stopping power even at v=8.0 a.u..

* E-mail: `ykanai@unc.edu`

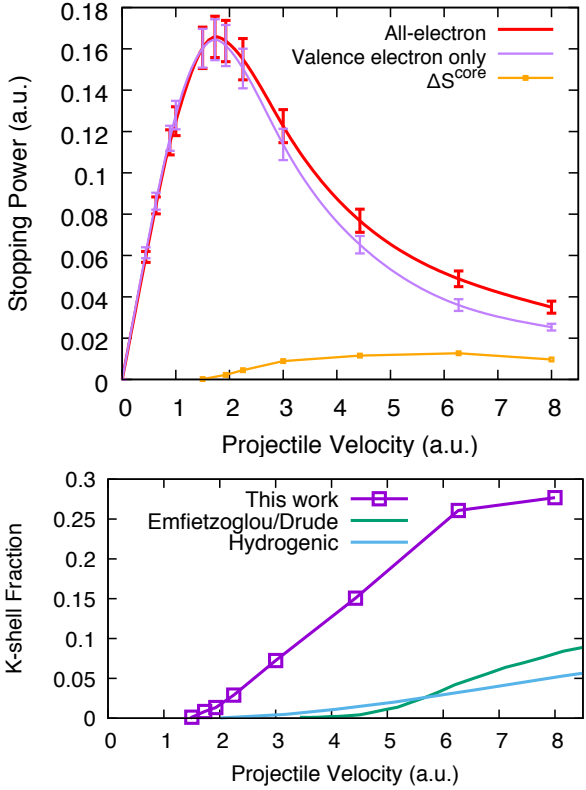

FIG 2 (Top) Contribution of the K-shell (oxygen 1s electrons) excitation to electronic stopping power curve, $\Delta S^{core}$, calculated as the difference between the all-electron and the valence-electron only results. (Bottom) The fraction of the K-shell contribution to the stopping power, in comparison to the Emfietzoglou/Drude model[11-12], and Hydrogenic generalized oscillator strength model[11, 45, 48-49].

As discussed above, K-shell core electron excitations not only provide an extra channel for the energy transfer, but they also influence valence electron excitations. To quantify this shake-up effect in the electronic stopping, we calculated the summed expectation value of the Kohn-Sham (KS) Hamiltonian for all the valence electron wavefunctions, $\sum \langle \varphi_i(t)|\hat{H}_{KS}|\varphi_i(t)\rangle$, in the simulations with and without the core electrons. The shake-up effect then can be quantified by obtaining the difference of this Hamiltonian expectation values for the valence electrons in the simulations with and without the K-shell core electrons. Figure 3 shows this energy difference as a function of the projectile proton displacement, averaged over all the 64 projectile paths. The shake-up effect contribution to the stopping power is obtained by calculating this expectation value change per unit distance of the projectile proton movement. At the high proton velocity of 8.0 a.u., the shake-up effect is responsible for 36 % of $\Delta S^{core}$ (i.e. 11 % of the stopping power). At the Bragg peak proton velocity of 1.73 a.u., 56 % of $\Delta S^{core}$ is due to the shake-up effect, but it is only <1% of the stopping power because K-shell core electrons are hardly excited at this peak velocity. For a very low velocity of 1.00 a.u., no shake-up effect is observed, and the difference between the all-electron and valence-electron-only calculations simply oscillates around zero in Figure 3. The K-shell core electron excitations have significant influence on the valence electron excitations at high velocities. Although having a separate correction for the core electron excitation is convenient in modeling[12, 23], it is not possible to take into account this intricate shake-up effect using such a model correction. This shake-up effect partly explains why using a separate K-shell correction underestimates the $\Delta S^{core}$ with respect to our first-principles result (see Figure 2 (bottom)).

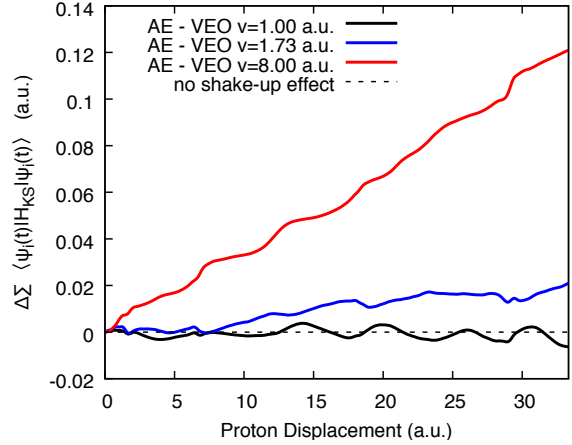

FIG 3 Difference of the summed Hamiltonian expectation value of the valence electrons for simulations with (all electrons - AE) and without K-shell core electrons (valence electron only - VEO) at the projectile proton velocity of 1.00 a.u., 1.73 (peak) and 8.00 a.u..

Having examined the K-shell core electron excitations and the importance of the shake-up effect, we now turn our attention to the spatial characteristics of the excited carriers in the electronic stopping process. The time-dependent Kohn-Sham (KS) wavefunctions can be projected onto the KS eigenstates of the equilibrium electrons to obtain the excited carrier distribution[17-18]. The projection onto the occupied and unoccupied eigenstates is used to calculate the hole and excited electron populations, respectively. All the occupied eigenstates and the unoccupied eigenstates up to 80 eV above the conduction band minimum are included in the projection, and the electronic states covered in this energy range account for greater than 95% of the total excited electrons. At the peak velocity of v=1.73 a.u., the average number of holes per water molecule is 0.0933, and only 0.003% ($3\times10^{-6}$ holes) are generated in the K-shell. At v=8.0 a.u, the average number of holes is significantly smaller, 0.0108, but approximately 1% ($1\times10^{-4}$ holes) of the holes are generated in the K-shell. Figure 4 shows the spatial distribution of the excited electrons and holes at v=8.0 a.u., as a function of the distance from each projectile proton path, averaged over all the projectile paths. A full width at half maximum (FWHM) of the distribution for the core holes is 0.40 Å, while a noticeably broader FWHM of 2.38 Å is observed for the hole distribution in the valence band states. The valence hole distribution shows two notable features: a localized region that corresponds to individual water molecules along the path and the distribution tail that derives from neighboring water molecules. This tail component gives the valence hole distribution an appreciable magnitude even at distances larger than 5 Å. On the other hand, the excited electron distribution is not so localized along the projectile proton path as shown in Figure 4, and the excited electron distribution decreases only by ~10% even at 5 Å away from the path. This indicates that individual water molecules are indeed ionized along the projectile path in the electronic stopping process, consistent with our earlier finding[17] and also with the established notion of proton radiation as ionizing radiation. The K-shell core electron excitations still contribute greatly to the stopping power even when only a small proportion of the excited electrons are excited from the K-shell core states because the core excitation energy is a few orders of magnitude greater than the valence excitation energy.

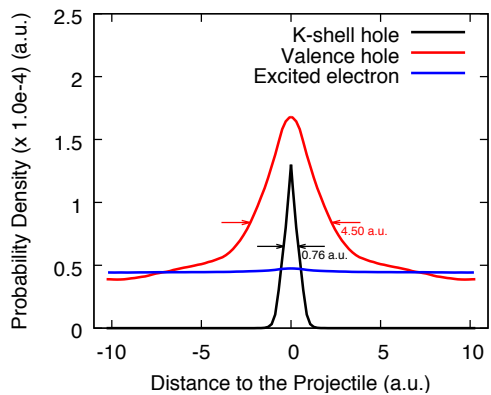

FIG 4 Ensemble-averaged distribution of holes and excited electrons as a function of the distance from the projectile proton path at the proton velocity of 8.0 a.u., the plot is made symmetric as a guide to the eye. The arrows indicate the FWHMs of the valence hole and O 1s hole distributions.

Developing a detailed understanding of the role of core electron excitations in liquid water under proton irradiation has become important largely due to the growing use of proton beams in radiation oncology. Using non-equilibrium simulations based on real-time time-dependent density functional theory, we accurately determined the electronic stopping power for protons in water from first principles, particularly focusing on the role of core electrons. The first-principles predicted stopping power shows significant differences to commonly-used perturbation theoretic models, such as the Bethe and Emfietzoglou models[12-13, 40]. The K-shell core electron excitation from water molecules' oxygen atoms was found to be crucial in determining the electronic stopping power curve beyond its maximum, being responsible for as much as one-third of the stopping power at the large proton velocity of 8.0 a.u. (the kinetic energy of 1.6 MeV). The core electron excitation significantly influences the valence electron excitation, in addition to providing an additional channel for the energy transfer. Such a cooperative phenomenon in the excitation is often refereed to as the shake-up effect[43], and this effect approximately accounts for as much as half of the contribution of the K-shell core electron excitation to the electronic stopping power at the high proton velocity of 8.0 a.u.. In the excitation process, the generated holes remain highly localized within a few angstroms around the irradiating proton path while electrons are excited away, indicative of ionizing radiation behavior. Despite their importance in contributing to the stopping power, the K-shell core electrons play a rather minor role in terms of the excitation density; only 1% of the holes is generated in the K-shell even at the large velocity of 8.0 a.u.. While X/γ-ray and proton radiations are both considered to be ionizing radiation and are usually treated on the same footing[19], our work revealed that the excitation/ionization behaviors involved are distinctly different.

## ACKNOWLEDGMENTS

The work is supported by the National Science Foundation under Grants No. CHE-1565714, No. DGE-1144081, and No. OAC-1740204. An award of computer time was provided by the Innovative and Novel Computational Impact on Theory and Experiment (INCITE) program. This research used resources of the Argonne Leadership Computing Facility, which is a DOE Office of Science User Facility supported under Contract DE-AC02–06CH11357.

\* E-mail: ykanai@unc.edu

* E-mail: ykanai@unc.edu

Supplemental Material for

# "K-shell Core Electron Excitation in Electronic Stopping of Protons in Water from First Principles"

Yi Yao, Dillon C. Yost, and Yosuke Kanai

Department of Chemistry, University of North Carolina at Chapel Hill

# Computational Details

## Planewave expansion and projectile proton

An important aspect of the accurate calculation of liquid water electronic stopping power for large velocities is the treatment of the proton projectile as a bare Coulomb potential. The bare Coulomb potential is significantly different than the pseudopotential, especially in the core region, leading to differences in electronic stopping power for large velocities. In Figure S1, the real-time time-dependent density functional theory (RT-TDDFT)-calculated stopping power data for v=8.0 a.u. is shown. When only valence electrons are explicitly treated using pseudopotentials for liquid water atoms and a pseudopotential is used also for the projectile proton, the electronic stopping power is 0.025 a.u.. When *all the electrons* are explicitly treated using pseudopotentials for liquid water atoms and a pseudopotential is used also for the projectile proton, the electronic stopping power is 0.028 a.u.. For this case, a very large planewave energy cutoff of 200 Ry is required to achieve the convergence. If *all the electrons* are explicitly treated using pseudopotentials for liquid water atoms and the projectile proton is represented exactly by using the bare Coulomb potential, the calculated stopping power is 0.035 a.u.. With the bare Coulomb potential proton projectile, the K-shell contribution to the stopping power is 3.5 times higher than the K-shell contribution calculated with the pseudopotential proton projectile.

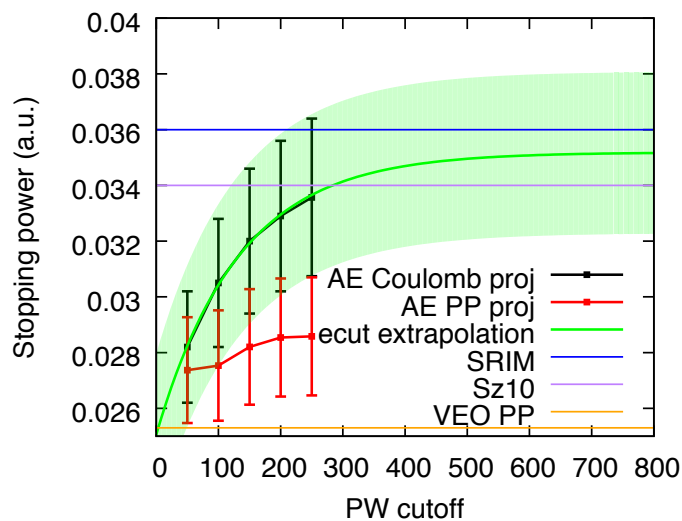

Figure S1: Convergence of the stopping power at the velocity of 8.0 a.u. AE Coulomb proj. (black) – All-electron calculation with bare Coulomb potential for the proton projectile, AE PP proj (red)– All-electron calculation with pseudopotential for the proton projectile, VEO PP proj (yellow) - Valence-electron only calculation with pseudopotential for the proton projectile. The planewave cutoff extrapolation (green) was performed with the formula of A*exp(-B)+C.

By projecting the time-dependent Kohn-Sham (TDKS) wavefunctions onto the ground state KS eigenfunctions, we acquire hole and excited electron distributions. As shown in Figure S2, with a proton projectile approximated by the pseudopotential, the hole population density of the K-shell is a small and smooth distribution. However, with the proton projectile represented by the Coulomb potential, the hole population for the K-shell is much larger and sharper.

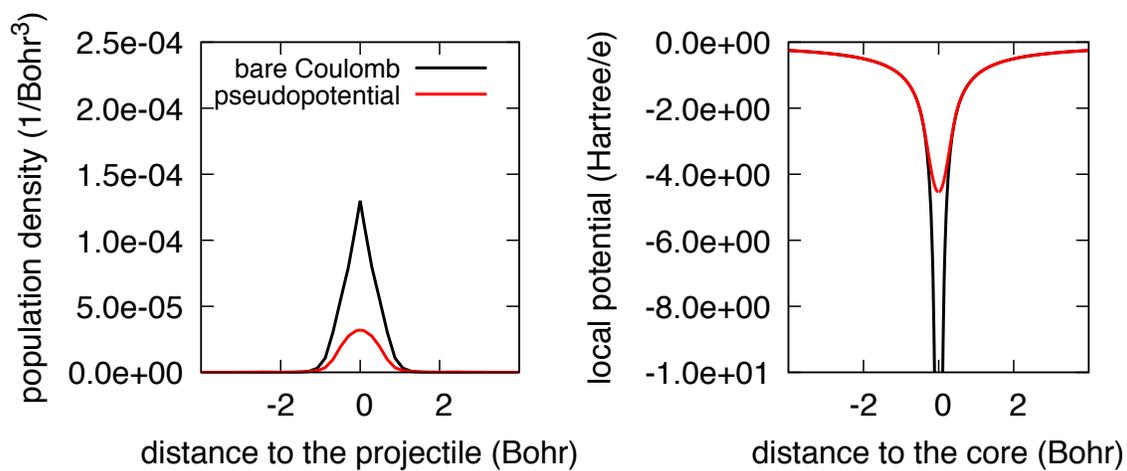

Figure S2: (Left) Population distribution for Oxygen 1s hole generated by bare the Coulomb proton projectile and the pseudopotential proton projectile. (Right) The local part of the pseudopotentials and bare Coulomb potential used for the projectile proton are shown.

**Projectile Proton Path Ensemble**

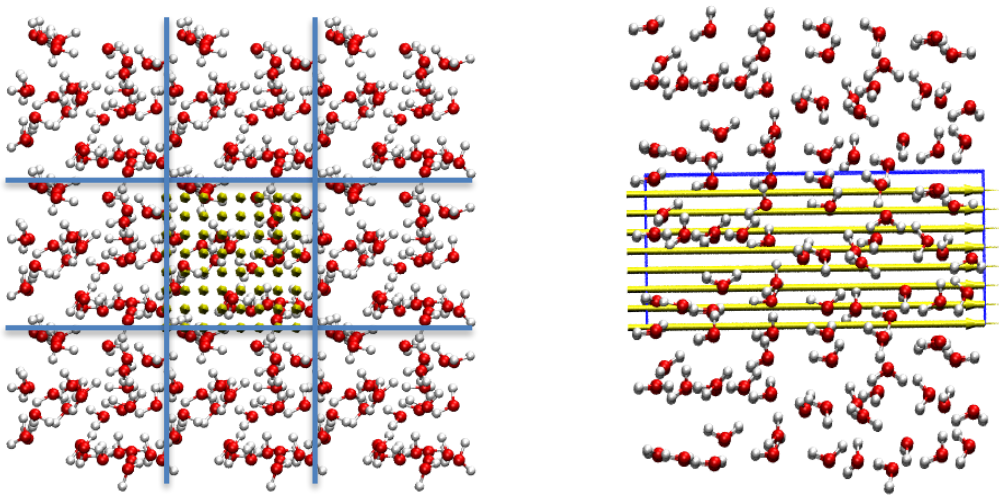

Figure S3. Front (left panel) and side (right panel) views of the sampled projectile paths for the stopping power calculations. The blue lines indicate the boundaries of the simulation cell that is periodically repeated.

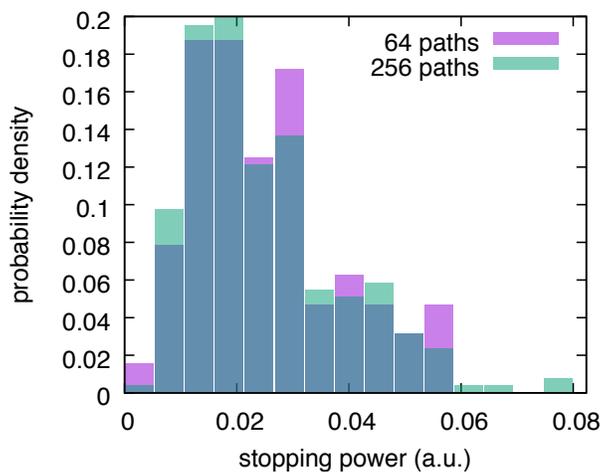

Figure S4. Path-dependent stopping power distribution for a proton with a velocity of 8.0 a.u.. for the 64-paths ensemble (purple), compared to a 256-path ensemble (green). Valence-electron only calculations with a pseudopotential for the proton projectile were used.

## Impact Parameter Sampling

Because the oxygen K-shell core electron excitation was found to be significant, having an accurate sampling of close (i.e. small) impact parameters is particularly important in the high velocity regime (e.g. v≈8 a.u.). For the electronic stopping power calculation, a classical ensemble average is taken over the 64 projectile (discussed above) proton trajectories in which the projectile proton is constrained to move along a straight path. In order to verify that deflection of the projectile proton by water molecules is negligible, Ehrenfest dynamics, using the RT-TDDFT forces, was performed for v=8 a.u.. Figure S5 shows how much the closest impact parameter changes for each of the 64 trajectories. As can be seen, whether we use the constrained or Ehrenfest dynamics, the trajectories are negligibly influenced in the sampling for even very small impact parameters. Because the average velocity remains close to v=8 a.u. at the end of the Ehrenfest dynamics (v=7.99992 a.u., to be specific), it is reasonable to calculate the stopping power for the Ehrenfest trajectories. This yields the stopping power value of 0.034979 (±0.002944) a.u., which is essentially identical to the stopping power of 0.034999 (±0.002943) for v=8 a.u., further validation that using the constrained trajectories has no effect on the impact parameter sampling and the stopping power calculation.

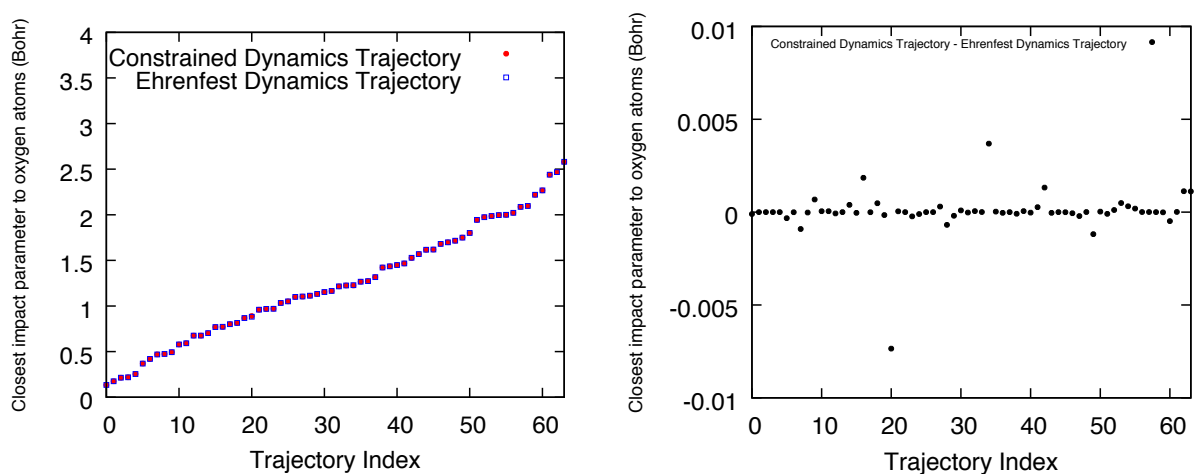

Figure S5. The closest impact parameter to oxygen atoms for all 64 trajectories. The trajectory indices are ordered by the closest impact parameter. (left) Absolute values for the Constrained Dynamics Trajectory and Ehrenfest Dynamics Trajectory. (right) The difference between Constrained Dynamics Trajectory and Ehrenfest Dynamics Trajectory.

**Simulation Cell Size**

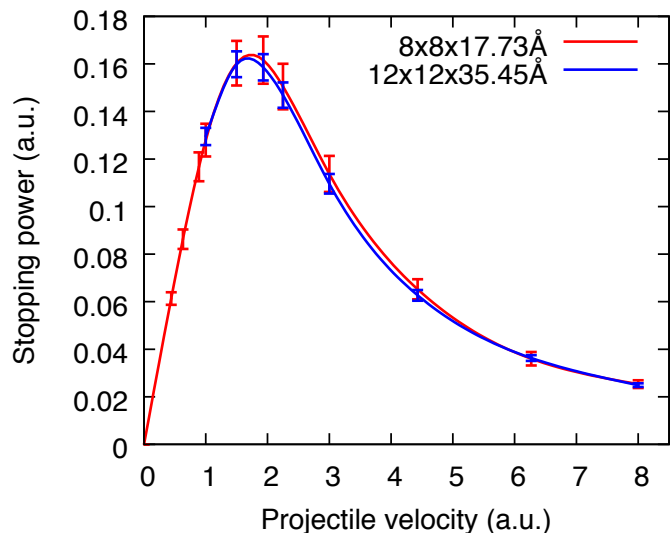

Figure S6. The stopping power curve calculated using simulation cell sizes of 8×8×17.73Å and of 12×12×35.45Å with periodic boundary conditions. Valence-electron only simulations with a pseudopotential for the proton projectile were used.

**Exchange Correlation (XC) Approximation Dependence: PBE GGA vs. SCAN meta-GGA**

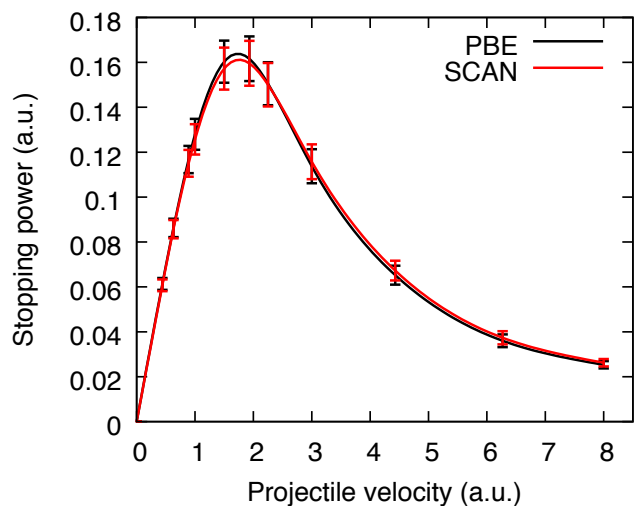

Figure S7: The stopping power curves calculated using the SCAN meta-GGA XC approximation and PBE GGA XC approximation Valence-electron only simulations with a pseudopotential for the proton projectile were used.

## Optical Excitation from 1s K-Shell in A Water Molecule

Because the plane-wave pseudopotential (PW-PP) formalism is generally not used for describing core electrons, we assessed the level of accuracy it can provide for modeling 1s K-shell optical excitation. The multi-projector approach by Hamann[1], enabled us to include the 1s electron in the calculation, as discussed in the main text. Using a RT-TDDFT simulation with the all-electron (AE) pseudopotential formalism, one is able to model the K-edge excitation from 1s electron of oxygen atoms in a water molecule while the valence-electron only (VEO) pseudopotential expectedly does not show the corresponding optical excitation. The K-edge of the absorption spectrum is in good agreement with the experimental measurement.

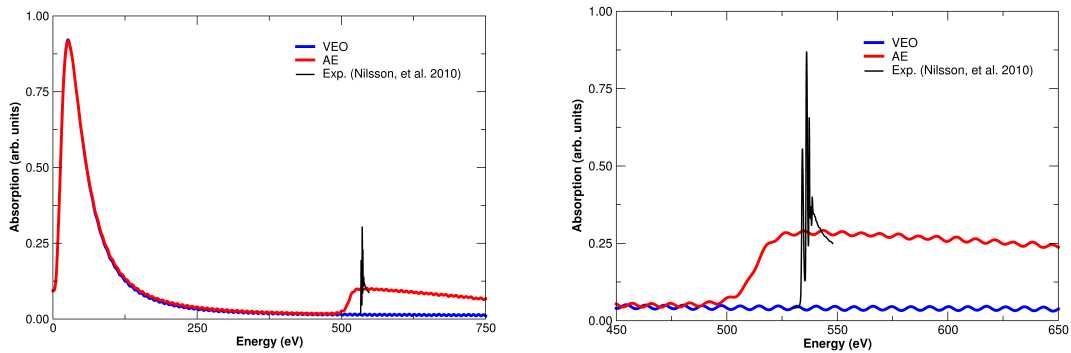

Figure S8. Single water molecule RT-TDDFT absorption spectra for the valence electron only (blue) and all-electron (red) cases. The all-electron RT-TDDFT K-shell core excitation onset is compared to the experimental X-ray absorption peak (black)[2].

# Excitation Density

Supplement to Figure 4 (in main text) including plots for the proton velocity of 1.73 a.u. (left panels) at the Bragg peak. Additionally, logarithmic scale plots are shown (bottom panels), as is necessary for the v=1.73 a.u. case because of the extremely small contribution to the hole density from the 1s K-shell of the oxygen atoms in water molecules.

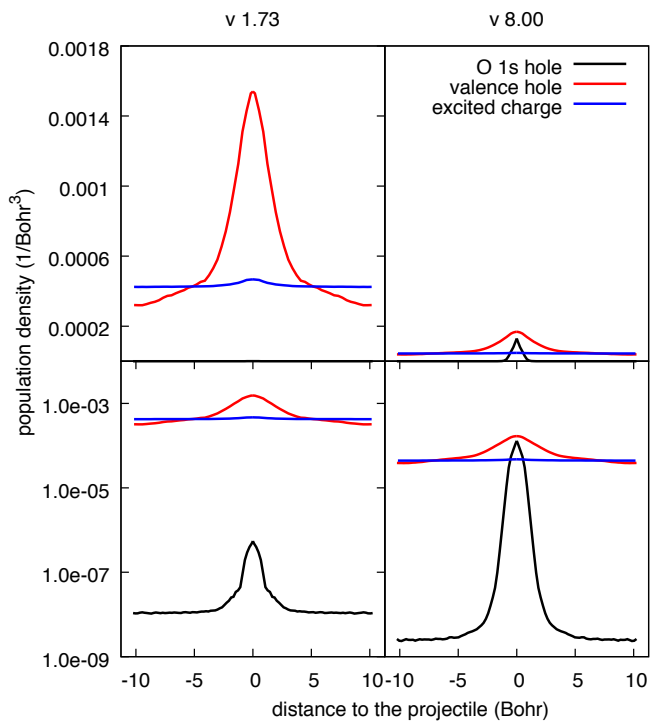

Figure S9. Ensemble-averaged distribution of holes and excited electrons as a function of the distance to the projectile proton path for the velocities of 1.73 (Bragg Peak) and 8.0 a.u., for convenience, the plot has been made symmetric. The logarithmic scale plots in the bottom reveal the small contribution to the total hole population from the oxygen 1s K-shell.

## Charge State of Projectile Proton

We calculated the charge state of the projectile ion from a summation of the electron density in its Voronoi cell[3]. The equilibrium electron density is first subtracted from the velocity-dependent, non-equilibrium electron density so that the electron density of liquid water does not contribute to the projectile ion charge[4].

We extract the charge state information from our first-principles calculation. The Schiwietz and Grande model[5] is also plotted for comparison. Note that the charge state is not an input for the RT-TDDFT simulations. Rather, it is a quantity that can be extracted from the induced electron density after the charge has reached a steady state on the projectile ion.

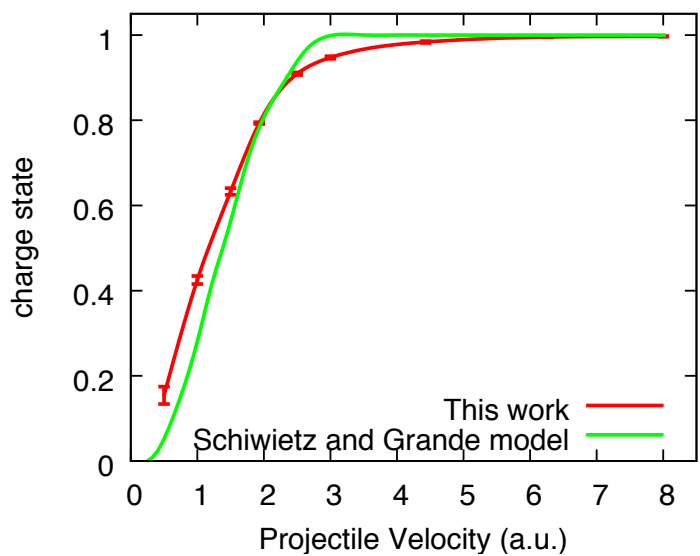

Figure S10. Mean steady-state charge for a proton projectile in liquid water as a function of the projectile ion velocity. Error bars represent the standard deviations of the distribution of the instantaneous charge state, which is calculated using the Voronoi partitioning[3] in the simulation. The empirical model for the projectile charge state by Schiwietz and Grande[5] is shown as green line for comparison.

## Comparison to earlier RT-TDDFT result and empirical/analytical models

We compare the electronic stopping power curve obtained using RT-TDDFT to our earlier work (Reeves, et al)[4] as well as to empirical/analytical models of Garcia-Molina[6], Penn[7-8], Ritchie[8-9], Ashley[8,10], and Emfietzoglou[8] discussed in our earlier work[4]. The work by Emfietzoglou, et al.[8] gives a detailed discussion of the models by Penn, Ritchie, and Ashley, along with the parameters used. The difference (within the statistical bars) between this first-principles work and our earlier first-principles work[4] derives from the inclusion of core electrons in our present work and also, to some extent, the more limited samplings of projectile proton paths in our earlier work[4].

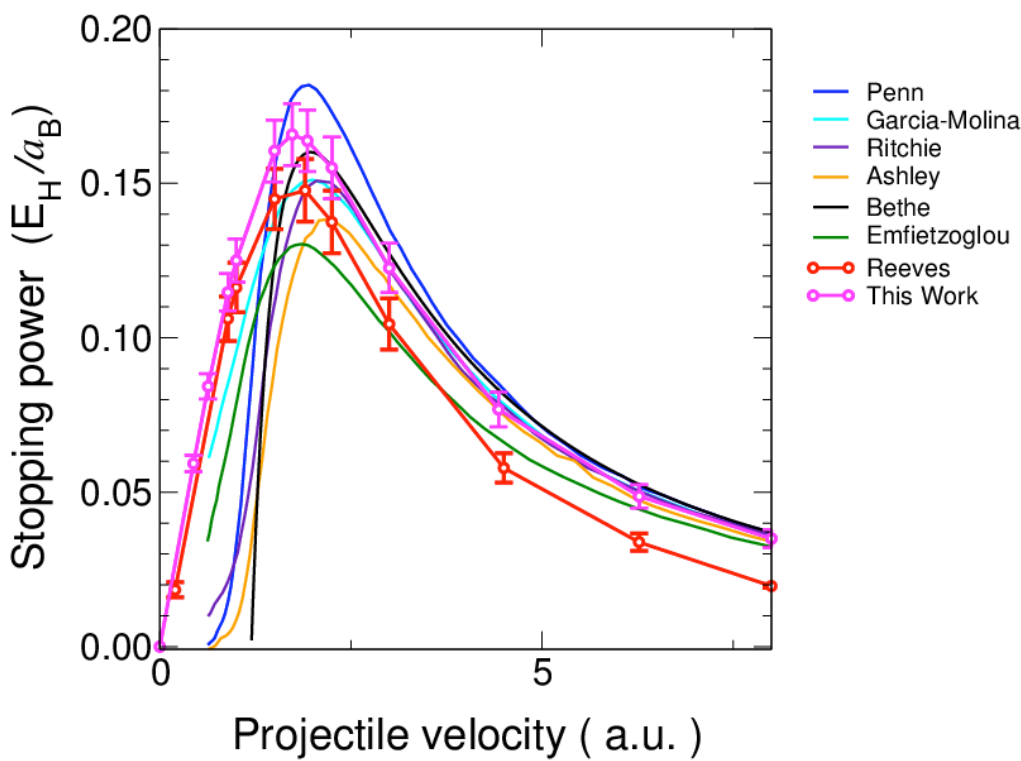

Figure S11. Electronic stopping power compared with our former work by Reeves, et al.[4] (red) and analytical models using various model dielectric functions by Garcia-Molina[6] (teal), Penn[7-8] (blue), Ritchie[8-9] (purple), Ashley[8, 10] (orange), and Emfietzoglou[8] (green). The Bethe formula result (black) using a mean excitation energy of I=78 eV (the value currently recommended by the International Commission on Radiation Units and Measurement[11]) is also plotted.

**Liquid water structure**

Cell vector a : 15.1178150222 0.0 0.0

Cell vector b : 0.0 15.1178150222 0.0

Cell vector c : 0.0 0.0 33.4969206901

```
atom  O1   O_species   11.2061   0.7370   24.1507
atom  H2   H_species   12.3210   1.3228   22.7523
atom  H3   H_species   10.6392  -1.0393   23.8106
atom  O4   O_species    5.5180   9.2030    6.7841
atom  H5   H_species    4.1763   8.3337    5.7826
atom  H6   H_species    4.7432  10.5069    7.9180
atom  O7   O_species   11.5651   2.4944    0.9827
atom  H8   H_species   13.1714   2.7590    0.0000
atom  H9   H_species   10.0911   2.8913   -0.1134
atom  O10  O_species    1.0016   3.8361    7.5400
atom  H11  H_species    0.7370   4.8188    5.9526
atom  H12  H_species    1.4929   5.0267    8.9384
atom  O13  O_species   13.1903  11.6974   18.2926
atom  H14  H_species   13.6249  12.0754   16.4973
atom  H15  H_species   12.3021  13.1903   19.0484
atom  O16  O_species    2.4377  14.2296    6.5951
atom  H17  H_species    2.1732  16.0816    6.6896
atom  H18  H_species    3.4204  13.8139    5.0456
atom  O19  O_species   10.6392   6.7463   15.5713
atom  H20  H_species    8.9573   7.4266   15.0422
atom  H21  H_species   11.7352   8.1825   16.1194
atom  O22  O_species    8.6361   1.7008   17.8957
atom  H23  H_species    8.5605   3.0614   19.2185
atom  H24  H_species    9.2975   2.4377   16.2705
atom  O25  O_species    3.6094   8.3526   22.8090
atom  H26  H_species    1.9275   7.5211   22.5633
atom  H27  H_species    4.7054   7.9935   21.3161
atom  O28  O_species   14.1541  11.6029   30.5569
atom  H29  H_species   15.7981  12.2454   29.8577
atom  H30  H_species   14.1541  11.8108   32.4466
atom  O31  O_species    0.7748   8.9384    9.5242
atom  H32  H_species   -0.6236   9.0140    8.2392
atom  H33  H_species    0.5102  10.3179   10.8092
atom  O34  O_species    8.7305   6.0471   21.2972
atom  H35  H_species    7.4077   6.7652   20.1634
atom  H36  H_species    7.8991   5.1212   22.7334
```

```
atom  O37  O_species   6.5574   2.0031    3.5338
atom  H38  H_species   6.3495   2.3622    1.6819
atom  H39  H_species   8.3715   2.0976    3.9873
atom  O40  O_species   6.7463   2.9102   31.7852
atom  H41  H_species   7.5589   4.1196   30.5758
atom  H42  H_species   6.3306   1.3039   30.8970
atom  O43  O_species  10.8281  10.6959   23.8106
atom  H44  H_species  11.3573   9.5998   25.2656
atom  H45  H_species  10.7714   9.6376   22.2421
atom  O46  O_species   4.2330  14.1163   17.9146
atom  H47  H_species   5.9148  14.9288   18.2737
atom  H48  H_species   3.5905  14.7210   16.2517
atom  O49  O_species   6.1227   9.2030    0.8504
atom  H50  H_species   5.4424   9.0707   -0.9071
atom  H51  H_species   4.6865   9.1652    2.0787
atom  O52  O_species  11.5840   1.5496    6.5385
atom  H53  H_species  11.6596   1.7008    4.6487
atom  H54  H_species  13.3226   1.7763    7.2566
atom  O55  O_species  11.4328   7.0487   29.3853
atom  H56  H_species  12.6045   8.4282   29.9522
atom  H57  H_species   9.6754   7.7290   29.2719
atom  O58  O_species   1.6819  14.6643    0.8693
atom  H59  H_species   3.3448  15.2123    1.5874
atom  H60  H_species   1.8519  14.4564   -1.0016
atom  O61  O_species   6.4440   8.0880   27.8924
atom  H62  H_species   6.6329   9.5809   26.7396
atom  H63  H_species   5.5936   6.6896   26.9475
atom  O64  O_species  11.1305  11.9242   11.9242
atom  H65  H_species  11.6029  11.1872   10.2423
atom  H66  H_species  10.2990  13.5871   11.6596
atom  O67  O_species   0.7748   4.0251   12.8312
atom  H68  H_species   0.5480   5.8959   12.8690
atom  H69  H_species  -0.6614   3.2314   11.8864
atom  O70  O_species   3.9117  13.2092   29.0262
atom  H71  H_species   3.7984  13.0769   27.1554
atom  H72  H_species   5.6881  12.8312   29.5931
atom  O73  O_species  14.0974   7.1810   20.6925
atom  H74  H_species  12.3777   6.4062   20.6169
atom  H75  H_species  14.0218   8.9573   20.0500
atom  O76  O_species   6.5196   8.3148   12.1887
atom  H77  H_species   7.9935   9.5242   12.1698
atom  H78  H_species   5.9148   8.0313   10.4313
atom  O79  O_species  10.2990   2.3433   12.3399
atom  H80  H_species  10.2234   4.0629   13.1147
atom  H81  H_species   9.5053   2.4000   10.6203
atom  O82  O_species  14.6643   1.6252   20.7114
```

```
atom H83   H_species   15.2879   3.4015   20.7681
atom H84   H_species   15.8737   0.5480   19.7287
atom O85   O_species    1.1338   9.0140    3.7984
atom H86   H_species    0.5480   7.9935    2.3244
atom H87   H_species    0.2457  10.6770    3.8172
atom O88   O_species    5.0078   6.8597   17.6689
atom H89   H_species    3.8172   5.4046   17.7634
atom H90   H_species    4.8566   7.6534   15.9493
atom O91   O_species    9.1652  13.2470   28.6861
atom H92   H_species   10.8281  13.1714   29.5931
atom H93   H_species    9.4108  12.7179   26.8908
atom O94   O_species    2.4755  14.2674   12.4344
atom H95   H_species    3.4582  14.1541   10.8281
atom H96   H_species    1.4929  15.8737   12.4722
atom O97   O_species   13.5304   2.1543   29.0829
atom H98   H_species   12.9068   3.9306   29.1585
atom H99   H_species   13.2281   1.4362   27.3632
atom O100  O_species    3.9306  13.7005   23.3759
atom H101  H_species    3.5527  11.8864   23.0547
atom H102  H_species    3.3259  14.7399   21.9208
atom O103  O_species   11.2061  11.1305    1.6630
atom H104  H_species    9.4297  10.6392    1.2472
atom H105  H_species   11.2061  12.2265    3.2125
atom O106  O_species    1.7385   6.2739   32.2387
atom H107  H_species    1.6252   5.4802   30.5191
atom H108  H_species    3.1936   5.5180   33.1836
atom O109  O_species    6.4062   3.1747   25.6247
atom H110  H_species    5.3479   1.6063   25.5680
atom H111  H_species    8.2203   2.7212   25.4168
atom O112  O_species   11.1872   9.0140    6.7463
atom H113  H_species   11.7919   9.1274    4.9511
atom H114  H_species    9.2975   9.0707    6.7841
```